\documentclass[a4paper,11pt]{article}
\usepackage{pos}
\usepackage[nameinlink]{cleveref}
\usepackage{subcaption}

\graphicspath{{figures/}}

\begin{document}
\title{Generative sampling with physics-informed kernels}
\author[a]{Friederike Ihssen}
\author*[a]{Renzo Kapust}
\author[a,b]{Jan M. Pawlowski}
\affiliation[a]{Institut f{\"u}r Theoretische Physik, Universit{\"a}t Heidelberg, Philosophenweg 16, 69120 Heidelberg, Germany}
\affiliation[b]{ExtreMe Matter Institute EMMI, GSI, Planckstr. 1, 64291 Darmstadt, Germany}
\emailAdd{kapust@thphys.uni-heidelberg.de}

\abstract{We construct a generative network for Monte-Carlo sampling in lattice field theories and beyond, for which the layerwise propagation is computed and optimised independently on each layer. The architecture uses physics-informed renormalisation group flows that provide access to the layerwise propagation step from one layer to the next in terms of a simple first order partial differential equation for the respective renormalisation group kernel through a given layer. 
Thus, it transforms the generative task into that of solving once the set of independent and linear differential equations for the kernels of the transformation. As these equations are analytically known, the kernels can be refined iteratively. This allows us to structurally tackle out-of-domain problems generally encountered in generative models and opens the path to further optimisation. We illustrate the practical feasibility of the architecture within simulations in scalar field theories. }

\FullConference{The $42^{\textrm{nd}}$ International Symposium on Lattice Field Theory (LATTICE 2025)\\
Tata Institute of Fundamental Research, Mumbai, India\\
2-8 November 2025
}

\maketitle

\section{Introduction}
\label{sec:Introduction}

Efficiently computing high-dimensional integrals poses a central task in lattice field theory calculations. This task is commonly conducted with Markov chain Monte Carlo approaches, where configurations are drawn from the respective Boltzmann distribution
\begin{align}
    p(\varphi) = \frac{1}{\mathcal{N}} e^{-\hat S(\varphi)} = e^{-S(\varphi)} \, \qquad \textrm{with} \qquad S(\varphi) = \hat S(\varphi) + \log \mathcal{N}\,,
    \label{eq:ProbabilityDistribution}
\end{align}
where we absorbed the normalisation $\mathcal{N}$ into the action $S(\varphi)$. However, as the phenomena of critical slowing down and topological freezing show, traditional sampling algorithms commonly lose their efficiency as one approaches the continuum limit of the theory~\cite{Wolff:1989wq, Alles:1996vn, DelDebbio:2004xh}. In these scenarios, the autocorrelation time of the Markov chains diverge, requiring prohibitively large amounts of computational resources. These problems may be overcome within generative models. These models aim to find transformations that move configurations from a theory with efficient sampling to configurations of the targeted theory, see e.g.~\cite{Luscher:2009eq, Rezende:2015ocs, Albergo:2019eim, Nicoli:2020njz}. 

However, generative Machine Learning models for Monte-Carlo simulations of lattice field theories face two out-of-domain (OOD) problems that hamper progress in this area. The most obvious one is related to the high dimensionality of the transformation between the two theories, which grows with the number of lattice sites and typically renders the learned distribution inaccurate. Roughly speaking, this originates in the fact that interpolation problems in high dimensions are equally ill-conditioned as extrapolation problems in low dimensions. Consequently, the computational costs of the final sampling steps with the true distribution scale badly with the dimension. This is a manifestation of the curse of dimensionality.

The second problem is a standard extrapolation problem and may be explained as follows: assume that the generative model has been trained with a given samples size related to a limited information content: the sample size is in one-to-one correspondence to the access to a specific set of moments or cumulants of the distribution with the orders $n < N_\textrm{max}$, where $N_\textrm{max}$ increases with the sample size but scales very badly. Further, sampling/training gives access to higher order moments/cumulants with $n > N_{\textrm{max}}$. However, the information carried by the higher order moments, or rather their irreducible parts, is not contained in the lower ones. Without any additional structure in the moments of the theory, this information is provided by the final accept-reject step with the true distribution which hence is increasingly expensive. We coin this OOD problem, the curse of limited information. Together, these two OOD problems lead to the bad scaling properties in the limit of large dimensions observed in many generative models.

Hence, the general situation asks for the construction of generative models or sampling algorithms for which the problems above are solved constructively rather than hoping for an underlying simplicity of the considered theory. Here, we describe a generative architecture based on \textit{physics-informed kernels} (PIKs) put forward in ~\cite{Ihssen:2025ybn}, based on physics-informed renormalisation group flows (PIRGs)~\cite{Ihssen:2024ihp}. 

The PIRG approach allows the construction of infinitesimal field transformations along the RG-time $t$ governed by a set of kernels $\dot\phi_t$ on the basis of the PIRG pair $(S_t, \dot\phi_t)$. The kernels can be directly computed from the respective RG equation for the pair. The key point of the PIRG in this approach is to analytically fix the action $S_t$ along this transformation. Then, the underlying renormalisation group equation is a simple linear differential equation for the kernel $\dot\phi_i$ at each time step. Importantly, this differential equation does not contain information from the previous times, thus completely decoupling the propagation steps from each other. This possible trivial parallelisation as well as the nature of the analytically known differential equation for the kernel at a given action is aimed to structurally resolve both of the OOD problems mentioned in the beginning. 

In the present work, we detail this approach based on PIKs in \Cref{sec:PIKsConcept}, where it is argued that PIKs allows for the construction of `truly' generative models. The conceptional developments are showcased within a zero-dimensional lattice field theory. In \Cref{sec:SummaryAndOutlook}, we conclude with a summary and an outlook of the many interesting further applications of the PIK-architecture.

\section{Sampling architecture with physics-informed kernels}
\label{sec:PIKsConcept}

Renormalisation group flows describe the scale dependence and general reparametrisations of a given statistical distribution. Both aspects are governed by the infinitesimal change of the distribution with the RG-time $t$. For the present purpose, we take the RG-time in a normalised time interval $t \in I$ with $I = [0,1]$. So, instead of the field $\varphi$ in \labelcref{eq:ProbabilityDistribution}, we consider fields $\phi_t$ that are related to $\varphi$ with a general non-linear transformation. This transformation comes with a change of the distribution $p_t(\phi)$. The respective differential equation is either derived for the normalised distribution $p_t(\phi)$ or its Laplace or Fourier transform, 
\begin{align}
    Z(J) = \int D\phi \,p_t(\phi)\, e^{\,\phi_i J_i} \,, \qquad \textrm{with} \qquad D\phi = \prod_{i\in \mathcal{D}} \int_{\mathcal{T}_t} d\phi_i\,,
\end{align}
where $\mathcal{D}$ denotes the lattice with lattice sites labeled by $i$ where the field value $\phi_i$ on that site takes values in $\mathcal{T}_t$. Here as for the remainder of this work, we use the Einstein sum convention. The most general scale and reparametrisation RG transformation is accommodated by the Wegner flow~\cite{Wegner:1974sla}, 
\begin{align}
    \frac{d p_t(\phi)}{dt} = - \frac{\partial}{\partial \phi_i} \left[\, \dot{\phi}_{t, i}(\phi) p_t(\phi) \, \right]\,. 
    \label{eq:WegnerEquation}
\end{align}
For related work in generative models and on its relation to optimal transport, see e.g.~\cite{Cotler:2023lem}. The field transformation related to \labelcref{eq:WegnerEquation} reads 
\begin{align}
    \frac{d \phi_{t, i}(\phi)}{dt} = \dot{\phi}_{t, i}(\phi)\,.
    \label{eq:WegnerFieldTransformation}
\end{align}
Here, the vector field $\dot\phi_t$ denotes the kernel of the transformation at the time $t$. The flow in \labelcref{eq:WegnerEquation} is a total derivative and hence, in the absence of boundary terms, leaves the total distribution $Z(J=0) = 1$ invariant. 

\begin{figure}[t]
    \centering
    \begin{minipage}[b]{0.49\textwidth}
        \centering
        \includegraphics[width=\textwidth]{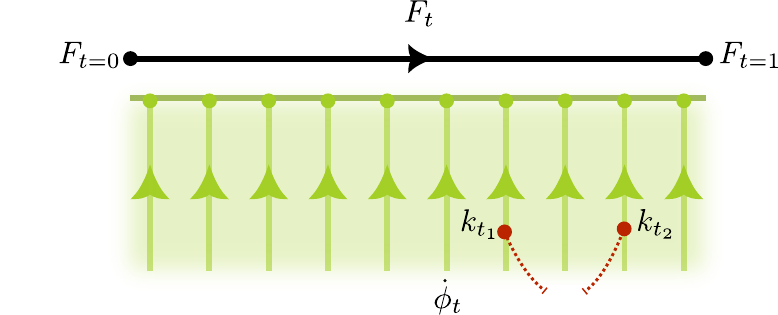}
        \vspace{0.01cm}
        \caption{Setup of the PIK-architecture, with fully accessible the path of $F_t$ (black) and kernels $\dot\phi_t$ (green) with completely independent parameters $k_t$ (red).}
        \label{fig:PIKSetup}
    \end{minipage}%
    \hfill%
    \begin{minipage}[b]{0.49\textwidth}
        \centering
        \includegraphics[width=\textwidth]{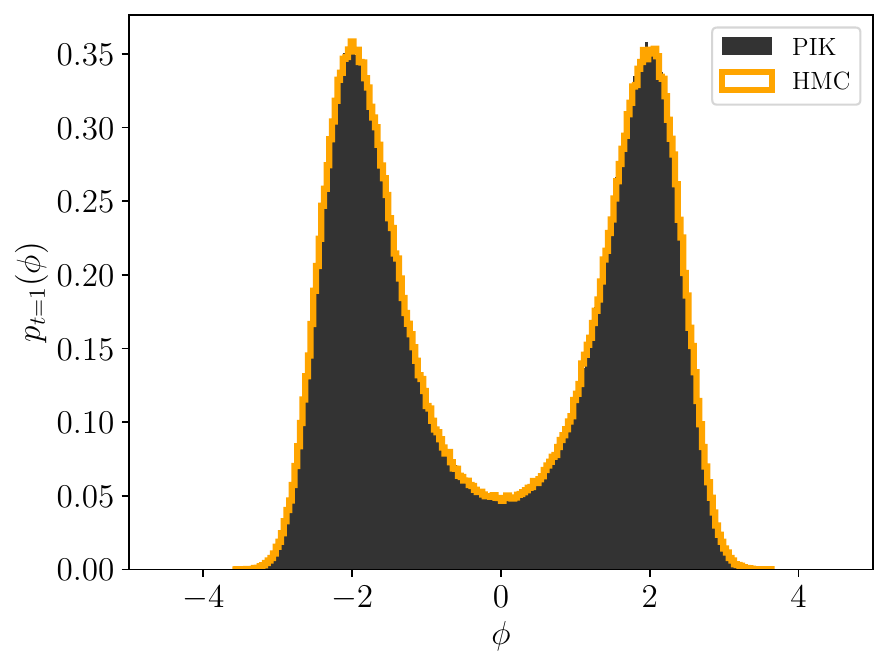}
        \caption{Comparison of the targeted (orange, HMC) and modelled (black, PIK) zero-dimensional $\phi^4$-theory distribution.}
        \label{fig:PIKHMCDistribution}
    \end{minipage}%
\end{figure}

A key idea underlying the physics-informed RG flows~\cite{Ihssen:2024ihp} is to exploit the full generality of the Wegner flow \labelcref{eq:WegnerEquation} for the distribution or related RG equations for other generating functions: Instead of solving the flows for the distribution $p_t(\phi)$, the rate function $\Gamma_t(\phi)$ or other generating functions $F_t$, we keep the coordinates $\phi$ general and solve the equations for pairs
\begin{align}
    \bigl(\,F_t(\phi), \dot\phi_t(\phi)\,\bigr)\,.
\end{align}
As mentioned in \Cref{sec:Introduction}, the key novel feature of the architecture proposed here is a structural property of the physics-informed RG flows: If the generating function $F_t \in \{p_t, \ldots\}$ is fully specified and known, the associated map $t\to F_t$ is completely independent of the previous ones,
\begin{align}
    \dot\phi_t(\phi) = \dot\phi_t(\phi; F_t)\,.
    \label{eq:KernelGivenByGenFunc}
\end{align}
Put differently, the kernel $\dot\phi_t$ of the map is \textit{physics-informed}. This perspective has also been used in~\cite{Albergo:2024trn}, where \labelcref{eq:KernelGivenByGenFunc} is solved within an architecture with a global loss function. In the present work, we use that \labelcref{eq:KernelGivenByGenFunc} leaves us with independent tasks of solving $\dot\phi_t$ for the whole time line from $t=0$ to $t=1$, which can be trivially parallelised. This is depicted in \Cref{fig:PIKSetup}, where $\dot\phi_t$ with parameters $k_t$ is computed vertically (green) to the evolution of $F_t$ (black).

Focusing on flows of the distribution $p_t(\phi) = e^{-S(\phi)}$, we consider the situation where we parametrised a path for the action $S_t$ which connects the theory of targeted action $S_1$ with a more tractable one, $S_0$. For the action, the Wegner equation can easily be rewritten as 
\begin{align}
    \frac{d S_t(\phi)}{dt} = \left[\, \frac{\partial}{\partial \phi_i} - \frac{\partial S_t(\phi)}{\partial \phi_i} \, \right] \dot\phi_{t, i}(\phi)\,.
    \label{eq:WegnerEquationForAction}
\end{align}
For a given $S_t$, this is a simple linear first order partial differential equation (PDE) for the physics-informed kernel $\dot\phi_t$ that has to be solved in order to transform the fields according to \labelcref{eq:WegnerFieldTransformation}.

As indicated by \labelcref{eq:ProbabilityDistribution}, the action given as $S_t(\phi) = \hat S_t(\phi) + \log \mathcal{N}_t$. Here, $\hat S_t(\phi)$ contains all field-dependent parts and $\log \mathcal{N}_t$ is field-independent and ensures normalisation. We parametrise
\begin{align}
    \hat S_t(\phi) = \sum_i c_{i,t } \mathcal{O}_{i,t}(\phi)\,,
    \label{eq:SHat}
\end{align}
with scale-dependent coefficients $c_{i,t}$ and a set of analytically known basis functions $\mathcal{O}_{i,t}$, which in principle can also be chosen time dependent. For the present work, we drop the operator's explicit time-dependence, for a short discussion see~\cite{Ihssen:2025ybn}. At the initial and final scale, $\{c_i, \mathcal{O}_i\}$ are chosen such that the action matches to the known base and target distributions. In between, the only constraint for the trajectory is its differentiability, and the ability to compute the field-derivatives $\partial S_t(\phi)/\partial \phi_i$. However, secondary constraints can be used for the optimisation of the sampling process, e.g. learning optimal base distributions.
 
The final ingredient in the PIK-architecture is the normalisation of the flowing distribution $p_t$. For PIKs, the flow of the normalisation $\mathcal{N}_t$ can be directly computed from that of $\hat S_t$ by
\begin{align}
    \frac{d \log \mathcal{N}_t}{dt} = - \int D\phi\,p_t(\phi)\,\frac{d \hat S_t(\phi)}{dt}\,.
    \label{eq:NormalisationFlow}
\end{align}
Actually computing (or approximating) this constant to the required precision is a complicated task which has to be avoided in practice. To this end, we rewrite the Wegner equation \labelcref{eq:WegnerEquationForAction} as 
\begin{align}
    \frac{d \log \mathcal{N}_t}{dt} = \left[\, \frac{\partial}{\partial \phi_i} - \frac{\partial \hat S_t(\phi)}{\partial \phi_i} \, \right] \dot\phi_{t, i}(\phi) - \frac{d \hat S_t(\phi)}{dt}\,.
    \label{eq:WegnerEquationFieldIndependence}
\end{align}
\Cref{eq:WegnerEquationFieldIndependence} allows us to compute the flow of the normalisation from the right hand side. Importantly, this circumvents the expensive sampling task \labelcref{eq:NormalisationFlow} and is key to the efficient sampling in the PIK-architecture. Furthermore, the kernel $\dot\phi$ can be computed by considering differences of Wegner equations for different field configurations $\phi$ and $\chi$. This leaves us with 
\begin{align}
    \frac{d \hat S_t(\phi)}{dt} - \frac{d \hat S_t(\chi)}{dt} = \left[\, \frac{\partial}{\partial \phi_i} - \frac{\partial \hat S_t(\phi)}{\partial \phi_i} \, \right] \dot\phi_t(\phi) - \left[\, \frac{\partial}{\partial \phi_i} - \frac{\partial \hat S_t(\phi)}{\partial \phi_i} \, \right] \dot\phi_t(\phi)\,\Big|_{\phi=\chi} =: \mathcal{L}_\chi \dot\phi_t(\phi)\,,
    \label{eq:WegnerEquationDifference}
\end{align}
where we introduced the linear operator $\mathcal{L}_\chi$ for notational convenience. Note that here all terms are analytically known and can be directly computed. Hence, one must not estimate or learn the change in the normalisation~\cite{Mate:2023toy,Albergo:2024trn} but obtains an efficient condition for the kernel $\dot\phi_t$.

In the following we consider solution strategies for the Wegner equation \labelcref{eq:WegnerEquation}, based on a parametrisation of the kernel $\dot\phi_t$ in some set of basis elements $\{K_{j, t}\}_{j=1}^M$,
\begin{align}
    \dot\phi_t(\phi) = \sum_{j=1}^M k_{j,t} K_{j,t}(\phi)\,,
    \label{eq:KernelParametrisation}
\end{align}
with the respective set of expansion coefficients $\{k_{j,t}\}_{j=1}^M$. With \labelcref{eq:WegnerEquationDifference}, this reduces the complicated problem of finding a transformation between two distributions to the problem of solving an independent set of linear systems of the form $A_t k_t = b_t$. Here, the matrix $A_t$ and vector $b_t$ are directly determined by the chosen basis $\{K_{j,t}\}_{j=1}^M$ and the parametrised $S_t$. 

We conclude this outline of the PIK-architecture with highlighting three key properties that follow from its structure and discuss them in the subsequent sections.
\begin{itemize}
    \item[(i)] \textit{Independent Kernels:} PIKs convert the global task of finding a precise invertible map from the base distribution $p_0$ to the final distribution $p_1$ within a deep network into independent tasks of computing the infinitesimal maps $\dot\phi_t(\phi;F_t)$ at each time step. For the pair $(S_t, \dot\phi_t)$ this amounts to solving \labelcref{eq:WegnerEquationForAction} for $\dot\phi_t$.
    \item[(ii)] \textit{OOD resolution:} The Wegner equation \labelcref{eq:WegnerEquationForAction} provides us with an error control of this map. This is also at the root of systematic improvements in the sampling process beyond the initial training time. In particular, curing the OOD problem with the PIRG pair ($S_t, \dot\phi_t$) boils down to locally improving the solution $\dot\phi_t$ to a known linear differential equation, whose coefficients are determined by $\hat S_t$.
    \item[(iii)] \textit{Optimisation:} The properties (i) and (ii) are relevant for the setup and solution of PIKs for a given pair $(F_t, \dot\phi_t$). The independent and analytically given structure of \labelcref{eq:WegnerEquationForAction} offers further freedom of optimisation for the pair, given external constraints. In particular, this includes a new kind of parameter conditional flow, systematic ways to improve inaccurate solutions of \labelcref{eq:WegnerEquationForAction} and the choice of the used generating function $F_t$.
\end{itemize}
%

\section{Physics-informed kernels at work}
\label{sec:PIKsWork}

We proceed with selected applications of the PIK-architecture to the commonly used benchmark
system of a zero-dimensional $\phi^4$-theory, illustrating the above three key properties.

\subsection{Independent kernels}
\label{sec:PIKsWorkIndependentKernels}

In order to illustrate the independent kernel structure of the PIK-architecture, we require a path for the action $S_t$ that connects a simple base distribution to the targeted one. For the zero-dimensional $\phi^4$-theory, an intuitive choice is given by 
\begin{align}
    \hat S_t(\phi) = \frac12 m_t^2 \phi^2 + \frac{1}{4} \lambda_t \phi^4\,\quad \textrm{with} \quad m_t^2 = m_0^2 + t \, (m_1^2 - m_0^2)\,, \quad \lambda_t = \lambda_0 + t \, (\lambda_1 - \lambda_0)\,.
\end{align}
Here, the mass $m_t$ and coupling $\lambda_t$ are chosen to be time-dependent, continuously connecting a base distribution with parameters $(m_0, \lambda_0)$ to another theory with parameters $(m_1, \lambda_1)$. More specifically, we use $(m_0^2 = 1,\, \lambda_0 = 0)$ and $(m_1^2=-2,\, \lambda_1=1/2)$, which transports a Gaußian into a multimodal interacting theory. For multimodal theories, PIKs circumvent the problem of mode-collapse~\cite{Hackett:2021idh, Chen:2022ytr, Nicoli:2023qsl} by enforcing a particular path $S_t$, which fixes the way the unimodal distribution is morphed into the multimodal one.

During the flow, the field $\phi_t$ is propagated in time according to \labelcref{eq:WegnerFieldTransformation}. This ODE in RG-time is solved numerically by a standard fourth order Runge-Kutta scheme with a step size of $\Delta t = 1/25$, leaving us with $N_t = 51$ independent kernels $\dot\phi_t$ to be determined. Each kernel is expressed in a basis with $K_{1,t} = \phi$ and $K_{j,t} = \sin(\omega_{j,t}\phi)$ for $j\geq 2$ and frequencies $\omega_{j,t}$, which respects the $\mathbb{Z}_2$ symmetry of the action~\cite{Koehler:2020efe,Gerdes:2022eve}. Using the efficient condition \labelcref{eq:WegnerEquationDifference}, the parameters of each kernel can be optimised independently without computing the change in the normalisation explicitly.

\Cref{fig:PIKHMCDistribution} shows a comparison of the targeted (orange) and modelled (black) distribution at the final time $t=1$. Here, the targeted distribution is obtained using $10^6$ samples from a standard hybrid Monte Carlo algorithm (HMC). The modelled distribution was generating by pushing $10^6$ samples from the initial Gaußian distribution through the PIK flow. We find an excellent agreement between both distributions.

\begin{figure}[t]
    \centering
    \begin{minipage}[t]{0.49\textwidth}
        \centering
        \includegraphics[width=\textwidth]{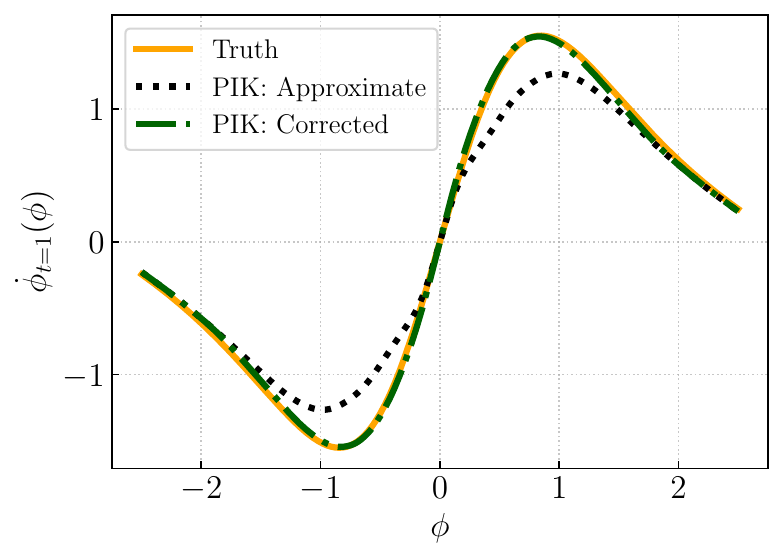}
        \caption{Comparison of the true (orange), the imperfect (black), and the corrected (green) kernel $\dot\phi_{t=1}$ for the zero-dimensional $\phi^4$-theory.}
        \label{fig:PIKCorrection}
    \end{minipage}%
    \hfill%
    \begin{minipage}[t]{0.49\textwidth}
        \centering
        \includegraphics[width=\textwidth]{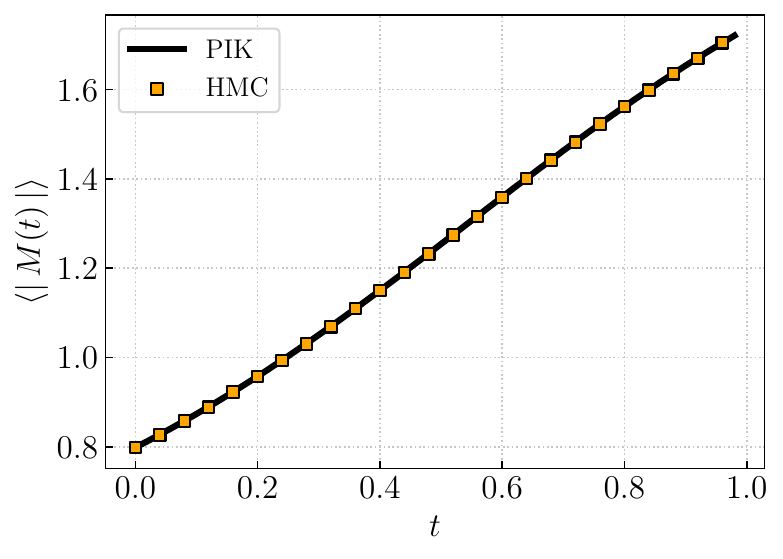}
        \caption{Magnetisation of the zero-dimensional $\phi^4$-theory computed with the PIK-architecture in a continuous manner (black) and HMC with individual simulations (orange).}
        \label{fig:PIKHMC_M_Comparison}
    \end{minipage}%
\end{figure}

\subsection{OOD resolution}\label{sec:PIKsWorkOODResolution}

Having showcased that PIKs have the capacity to compute each kernel $\dot\phi_t$ independently without the requirement of computing the change in the normalisation, we know turn ot the property (ii) \textit{OOD resolution}. Instead of the explicit volume scaling, which is part of ongoing research, we illustrate the systematic improveability of the PIK sampling process after its initial setup. This is naturally enabled by the fact that we have analytical access to the Wegner equation \labelcref{eq:WegnerEquationForAction,eq:WegnerEquationDifference} that has to be satisfied by each kernel individually. It is further boosted by the fact that the Wegner equation is linear in $\dot\phi_t$, making iterative improvements directly accessible.

In order to mimic the OOD problem of the curse of limited information, we purposefully only compute the final kernel on a dataset limited to $50k$ samples. This approximate kernel is referred to as $\dot\phi_t^{(0)}$. This kernel (black) together with the true solution (orange) is depicted in \Cref{fig:PIKCorrection}. Note, that also for other continuous generative models, see e.g.~\cite{Chen:2018wjc,Gerdes:2022eve,Bulgarelli:2024brv,Aarts:2026zzr}, the Wegner equation at the last time step must be solved exactly \labelcref{eq:WegnerEquationForAction} in order to be truly generative. 

In order to systematically improve the sampling, we must improve the solution of the kernel $\dot\phi_t^{(0)}$ such that it provides a more accurate solution of the Wegner equation. There are many ways to achieve this goal. Here, we use the linear structure of \labelcref{eq:WegnerEquationDifference}. The residual of \labelcref{eq:WegnerEquationDifference} with the approximate kernel $\dot\phi_t^{(0)}$ is then readily computed as 
\begin{align}
    r^{(0)}_\chi(\phi) = \frac{d \hat S_t(\phi)}{dt} - \frac{d \hat S_t(\chi)}{dt} - \mathcal{L}_\chi \dot\phi_t^{(0)}(\phi)\,.
    \label{eq:Residual}
\end{align}
Naturally, a correction $\Delta\dot\phi^{(0)}_t$, that satisfies $\mathcal{L}_\chi \Delta\dot\phi^{(0)}_t(\phi) = r^{(0)}_\chi(\phi)$, will lead to a corrected kernel $\dot\phi_t^{(1)} = \dot\phi_t^{(0)} + \Delta\dot\phi^{(0)}_t$ that solves the Wegner equation \labelcref{eq:WegnerEquationDifference} accurately. This procedure can also be iterated as the Wegner equation is linear in $\dot\phi_t$, hence the superscripts, indicating the first iteration. 

In order not to spoil the well-fitted regions of the approximate kernel, we solve for the correction with a collocation approach using compactly supported radial basis functions (RBFs), more concretely antisymmetrised Wendland functions. 
As the RBFs have a compact support, the resultant linear system is sparse and readily solved.

In \Cref{fig:PIKCorrection}, we show the corrected kernel $\dot\phi_t^{(1)}$ in green. We find that the correction indeed leads to a kernel that is very close to the true one and corrects the approximate kernel in the regions with the largest deviations. While the correction mechanisms can be readily improved, this example illustrates the general systematic workflow of PIKs from (a) an (imprecise) global solution to (b) a determination of OOD regions during sampling or using information of the residual to (c) a residual-driven correction of the kernel. This is natural for PDE approaches but the proposed method is to our knowledge novel in the context of generative models.

\subsection{Optimisation}
\label{sec:PIKsWorkOptimisation}

The above sections showcased the setting, computation, and systematic improvement within the PIK-architecture. In this section, we comment on the property (iii) \textit{Optimisation} of PIKs. As outlined in \Cref{sec:PIKsConcept}, there are various ways to optimise the pair PIRG pair. These reach from the choice of the generating function to optimal paths for the same, improved base distributions and to the optimisation of the solution method of the flow equation. Here, we illustrate one such optimisation at the example of a parameter-conditional flow.

With PIKs we have direct control over the distribution at each time step. Here, we use this freedom to enforce the same functional form of the action at each time step. The only change of  the distribution in the flow are the parameters ($m^2, \lambda)$. This allows us to make PIKs parameter conditional without inflating the network, which is the usual strategy to make the parameter of the generative model conditional~\cite{Gerdes:2022eve,Singha:2022icw}. Moreover, this directly allows us to compute observables continuously without sampling the configurations for each parameter.

We exemplify this PIK approach in \Cref{fig:PIKHMC_M_Comparison} for the magnetisation
\begin{align}
    \langle | M(t) | \rangle = \int D\phi\,p_t(\phi)\,|\phi|\,,
\end{align}
For the PIK-architecture (black), this observable can be computed by pushing $10^6$ Gaußian samples from the initial distribution through the flow and computing the magnetisation at each intermediate step. For comparison, we also show the same observable computed with individual HMC simulations (orange), where each point corresponds to a separate simulation.

\section{Summary and outlook}\label{sec:SummaryAndOutlook}

We have  described physics-informed kernels (PIKs) as a new architecture for generative models as constructed in \cite{Ihssen:2025ybn}. The PIK-architecture aims at solving OOD problems in generative models as well as implementing a systematic error control. At its heart, PIKs reduce the task of finding transformations between two distributions into a decoupled chain of layerwise transformations that are given by an analytically known renormalisation group equation. For the distribution itself, this is the Wegner equation \labelcref{eq:WegnerEquation}, for the rate function it is the generalised flow equation~\cite{Pawlowski:2005xe}, and similar ones exist for other generating functions. Viewed as PDEs for the kernels, these equations boil down to independent linear differential equations. Importantly, their solution can be systematically corrected at each time step without requiring additional steps in the Markov chain, which is different to most prominent generative models like normalising flows or diffusion models, see e.g.~\cite{Chen:2018wjc,Albergo:2019eim,Song:2020hus,Gerdes:2022eve,Lipman:2023hef,Bulgarelli:2024brv,Singha:2025lsd,Aarts:2026zzr}. As a paradigmatic example, we have considered PIKs for the distribution with completely determined action paths $S_t(\phi)$ that lead to simple Wegner equations.

This setup paves the way for many interesting applications and improvements of the PIK-architecture. The first prominent direction is the application of PIKs to higher-dimensional field theories as well as the extension to field theories with gauge or fermionic degrees of freedom. Moreover, as the PIK-architecture enables general transformations of the measure it is tailor-made to tackle sign problems, where the action becomes complex. This is particularly relevant for real-time simulations and finite density field theories and is part of a forthcoming work~\cite{PIK2026}. Furthermore, the PIK-architecture enables further optimisations of the sampling process, e.g. by computing optimal paths for $S_t$, including optimal base distributions. Lastly, going beyond sampling, the PIK-architecture is also envisioned to infer the action or Hamiltonian in a symbolic way from data, which is also part of ongoing research.

\acknowledgments{
We thank Gert Aarts, Kenji Fukushima, Sander Hummerich, Thore Kolja Joswig, Ullrich Köthe, Timoteo Lee, Manfred Salmhofer, Robert Scheichl, Lingxiao Wang and Kai Zhou for discussions and collaborations on related subjects. We acknowledge the DM-LFT Collaboration for stimulating discussions and continuous support throughout this work. It is funded by the Deutsche Forschungsgemeinschaft (DFG, German Research Foundation) under Germany’s Excellence Strategy EXC 2181/1 - 390900948 (the Heidelberg STRUCTURES Excellence Cluster). We also acknowledge support by the
state of Baden-Württemberg through bwHPC.}

\bibliographystyle{JHEP}
\bibliography{../refs.bib}

\end{document}